%% file: main.tex
\documentclass[conference]{IEEEtran}
\IEEEoverridecommandlockouts
\usepackage[T1]{fontenc}
\IEEEoverridecommandlockouts
\usepackage{cite,cuted}
\usepackage{amsmath,amssymb,amsfonts}
\usepackage{algorithm,algorithmic}
\usepackage{graphicx}
\usepackage{textcomp,amsfonts}
\usepackage{xcolor}
\usepackage{tikz,float}
  \newlength\fheight
\newlength\fwidth
\usepackage{tkz-euclide}
\usepackage{hyperref}
\def\BibTeX{{\rm B\kern-.05em{\sc i\kern-.025em b}\kern-.08em
    T\kern-.1667em\lower.7ex\hbox{E}\kern-.125emX}}
    \usepackage{pgfplots} 
\usepackage{pgfgantt}
\usepackage{pdflscape}
\pgfplotsset{compat=newest} 
\pgfplotsset{plot coordinates/math parser=false}
\pgfplotsset{every  tick/.style={black,},ylabel style={font=\tiny},xlabel style={font=\tiny},tick label style={font=\tiny},legend style= {font=\scriptsize},
minor x tick num=1,minor y tick num=1,xminorticks=true,yminorticks=true,}  
\usepackage{caption}
\newtheorem{theorem}{Theorem}
\begin{document}
\title{Adaptive Gradient Search Beamforming for Full-Duplex mmWave MIMO Systems }

\author{\IEEEauthorblockN{Elyes Balti}
\IEEEauthorblockA{
\textit{Wireless Networking and Communications Group} \\
%\textit{Department of Electrical and Computer Engineering} \\
\textit{The University of Texas at Austin}\\
Austin, TX 78712, USA \\
ebalti@utexas.edu}
%\and
%\IEEEauthorblockN{\textsuperscript{} Neji Mensi}
%\IEEEauthorblockA{\textit{Department of Electrical Engineering and Computer Science} \\
%\textit{Howard University}\\
%Washington, DC 20059, USA \\
%neji.mensi@bison.howard.edu }
}
\maketitle
\begin{abstract}
In this work, we present a framework analysis of full-duplex (FD) systems for Millimeter Wave (mmWave) analog architecture. Given that FD systems can double the ergodic capacity, such systems experience large losses caused by the loopback self-interference (SI). In addition, systems with analog architecture also suffer from other forms of losses mainly incurred by the constant amplitude (CA) constraint. For this purpose, we propose the projected Gradient Ascent algorithm to maximize the sum rate under the unit-norm and CA constraints. Unlike previous works, our approach achieves the best spectral efficiency while minimizing the losses incurred by the CA constraint. We also consider an adaptive step size to compensate for the perturbations that may affect the cost function during the optimization. The results will show that the proposed algorithm converges to the same optimal value for different initializations while the number of iterations required for the convergence changes for each case. In this context, we primarily consider the gradient search method for a two-nodes FD systems and then we extend the analysis for a dual-hop FD relaying systems. Finally, we evaluate the robustness of our method in terms of rate and outage probability and compare with previous approaches.
\end{abstract}

\begin{IEEEkeywords}
Gradient search, full-duplex, constant amplitude constraint, millimeter wave, MIMO. 
\end{IEEEkeywords}
\IEEEpeerreviewmaketitle

\input{introduction.tex}
\input{systemmodel.tex}

\input{twonode.tex}
\input{relay.tex}
\input{numerical.tex}

\input{conclusion.tex}

\bibliographystyle{IEEEtran}
\bibliography{main}
\end{document}

%% file: introduction.tex
\section{Introduction}
FD systems have been intensively investigated in the literature due to its potential to double the spectral efficiency compared to half-duplex (HD) systems \cite{e1,e2,e3,e4,e5,e6,e7,eT,asym}. Hence, combining the mmWave technology and FD systems can provide ultra higher spectral efficiency. Such mmWave based FD systems gained enormous attention as it can be applied in wide ranges of applications requiring large bandwidth. For example, FD mmWave systems can be a practical solution for autonomous driving, platooning, vehicular clouding/security and advanced driving assistance systems (ADAS) \cite{maalej,neji1,neji2,surv,rheath,5gnr}. Such applications require low latency, large bandwidth and efficient link budget to exchange big data and processing \cite{v2v}. It is true that FD systems are rate-efficient, however, they suffer from the loopback self-interference (SI) caused by the transmission and reception at the same frequency/time resources. Without interference cancellation, related works have shown that this loopback SI can significantly reduce the spectral efficiency, making FD systems impractical. In this context, related works proposed various techniques to cancel the SI. Conventional approaches are hardware based such as antennas array structures (separation, isolation, polarization, directivity), analog and digital cancelling circuits \cite{sub6}. For example, antennas isolation can achieve about 15 dB of SI reduction \cite{isol,19}, while hybrid analog-digital circuits can suppress up to 70 dB of SI leaving a residual interference of 3 dB higher than the noise floor and resulting in roughly 67-76 $\%$ of rate improvement compared to HD systems \cite{analog,digital}. However, recent works have focused more on spatial beamforming based techniques as they are less susceptible to the hardware constraints. Literature proposed various beamforming approaches such as angle seach, beam steering, minimum mean square error (MMSE), Zero-Forcing (ZF) max power design \cite{14,unconst}. Performances of these algorithms have been discussed and evaluated, however, they experienced severe losses incurred by the constant amplitude (CA) constraint. In fact, once the beamformers are optimally designed for each algorithm, the CA constraint violates the SI cancellation constraints resulting in huge rate losses. To address this limitation, \cite{balti2020modified} proposed a modified variant of ZF max power to efficiently introduce the CA constraint while minimizing the losses. 
%\subsection{Contribution}
In this context, we propose a robust gradient search beamforming to cancel the SI and minimize the rate losses incurred by the CA constraint. To the best of our knowledge, our proposed algorithm provides the smallest gap between the full-digital and analog beamforming with the CA constraint. The analysis of this paper follows these steps:
\begin{enumerate}
    \item Present a detailed analysis about the systems architectures and the channels models.
    \item Formulate the problems and derive the gradients.
    \item Provide the Gradient search algorithm to solve the problem for the two-nodes FD network and then extend the method for the dual-hop FD relaying system.
\end{enumerate}
This paper is organized as follows: Section II describes the system and channels models while Section III provides the analysis of the two-nodes full-duplex network along with the problem formulation and Gradient Ascent algorithm. Section IV extended the results for an application of dual-hop full-duplex relaying systems where we formulated the spectral efficiency along with the outage and error probabilities. Section V provides the numerical results while the concluding remarks and future direction are reported in Section VI.
  

%% file: systemmodel.tex
\section{System Model}
In this work, we adopt the geometric channel model based on clusters and rays as follows %\cite[Eq.~(4)]{sparse}
\begin{equation}\label{channel}
\textbf{H} = \sqrt{\frac{N_{\text{RX}}N_{\text{TX}}}{N_{\text{cl}}N_{\text{ray}}}} \sum_{k=1}^{N_{\text{cl}}}\sum_{\ell=1}^{N_{\text{ray}}}\alpha_{k,\ell}\text{a}_{\text{RX}}(\phi_{k,\ell},\theta_{k,\ell})  \text{a}^*_{\text{TX}}(\phi_{k,\ell},\theta_{k,\ell})  
\end{equation}
where $N_{\text{cl}}$ and $N_{\text{ray}}$ are the number of clusters and rays, respectively, $\alpha_{k,\ell}$ is the complex gain the $\ell^{\text{th}}$ ray in the $k^{\text{th}}$ cluster, and $\text{a}_{\text{TX}}(\phi_{k,\ell},\theta_{k,\ell})$ and $\text{a}_{\text{RX}}(\phi_{k,\ell},\theta_{k,\ell})$ are the array steering and response vectors at transmitter (TX) and receiver (RX), respectively, evaluated at the angles of departure (AoD) and the angles of arrival (AoA). 
\begin{figure}[H]
\label{position}
    \centering
    \includegraphics{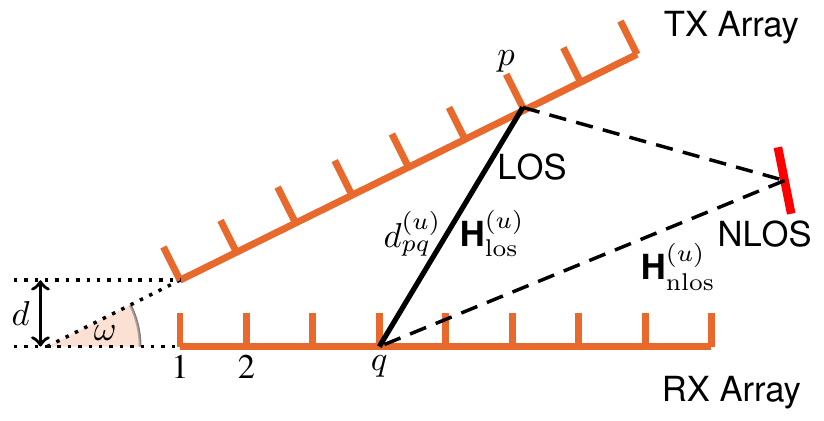}
    \caption{Relative position of TX and RX arrays at FD node $u$. Given that the TX and RX arrays are collocated, the far-field assumption that the signal impinges on the antenna
array as a planar wave does not hold. Instead, for FD transceivers, it is more suitable to assume that the signal impinges on the array as a spherical wave for the near-field LOS channel. }
\end{figure}
The SI leakage is decomposed into a line of sight (LOS) component modeled by $\textbf{H}_{\text{los}}$ and a non-line of sight (NLOS) leakage described by $\textbf{H}_{\text{nlos}}$ which is a random complex Gaussian matrix. The LOS SI leakage matrix can be written as 
\begin{equation}
 [\textbf{H}_{\text{los}}]_{pq} = \frac{1}{d_{pq}}e^{-j2\pi\frac{d_{pq}}{\lambda}}    
\end{equation}
where $d_{pq}$ is the distance between the $p$-th antenna of the TX array and $q$-th antenna of the RX array. The aggregate SI channel can be obtained by
\begin{equation}\label{channelSI}
\textbf{H}_{\text{SI}} = \sqrt{\frac{\kappa}{\kappa+1}}\textbf{H}_{\text{los}} + \sqrt{\frac{1}{\kappa+1}}\textbf{H}_{\text{nlos}}    
\end{equation}
where $\kappa$ is the Rician factor.

%% file: twonode.tex
\section{Two-Nodes Full-Duplex Network Analysis}
\begin{figure}[H]
    \centering
    \includegraphics[scale=.7]{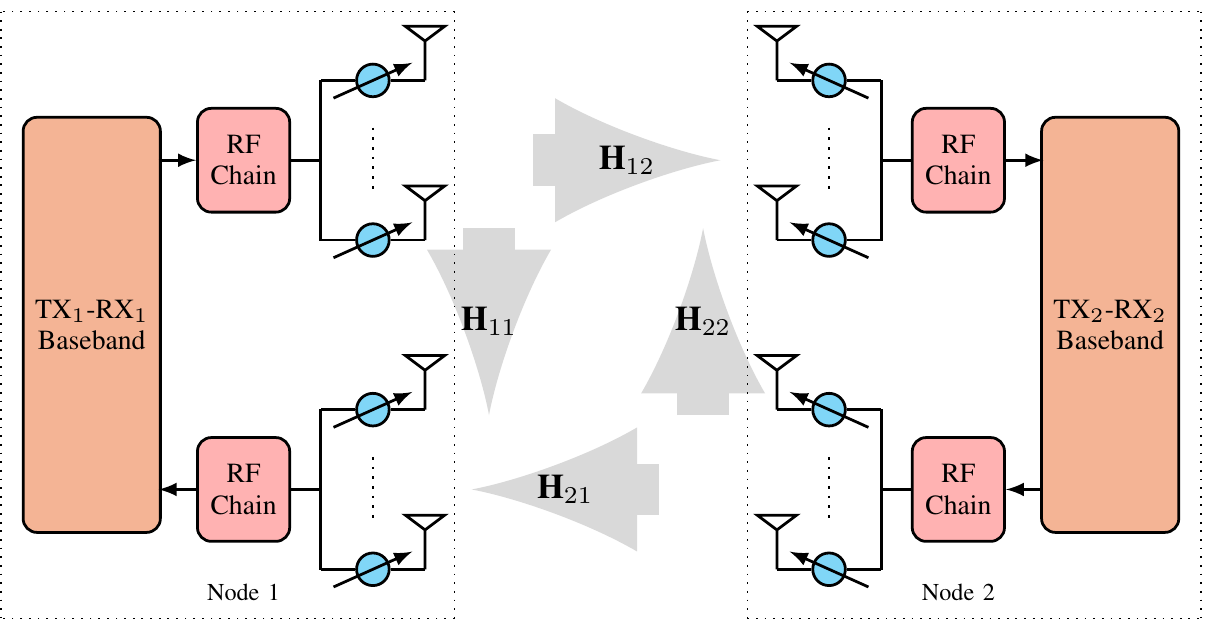}
    \caption{Architecture of a two-nodes FD network. The channels $\textbf{H}_{12}$ and $\textbf{H}_{21}$ are the geometric channels defined by Eq.~(\ref{channel}) while the SI channels $\textbf{H}_{11}$ and $\textbf{H}_{22}$ are defined by Eq.~(\ref{channelSI}).}
\label{system}
\end{figure}

Fig.~\ref{system} presents an analog architecture of a mmWave two-nodes FD communicating with each others. Such structure supports single stream with one RF chain and each node consists of a transmit and receive antennas array.
The received signals at the two nodes are expressed by
\begin{equation}
y_1 = \textbf{w}_1^*(\sqrt{\rho_2}\bold{H}_{21}\textbf{f}_2 s_2 + \sqrt{\tau_1}\textbf{H}_{11}\textbf{f}_1s_1 + \textbf{n}_1)    
\end{equation}
\begin{equation}
y_2 = \textbf{w}_2^*(\sqrt{\rho_1}\textbf{H}_{12}\textbf{f}_1 s_1 + \sqrt{\tau_2}\textbf{H}_{22}\textbf{f}_2s_2 + \textbf{n}_2)    
\end{equation}
where $s_u$ is the information symbol sent by node $u$, $\rho_u$ is the average transmit power at node $u$, $\tau_u$ is the SI power leakage at node $u$, $\bold{f}_u$ is the precoder applied at the transmit node $u$, $\bold{w}_u$ is the combiner at the receive node $u$, and $\bold{n}_u$ is the zero-mean AWGN noise with variance $\sigma_u^2\bold{I}_{N_{r,u}}$. 
Consequently, the sum rate is given by
\begin{equation}\label{rate}
\begin{split}
\mathcal{I} =& \log\left(1 + \frac{\rho_1|\textbf{w}_1^*\bold{H}_{21}\textbf{f}_2 |^2}{\sigma_1^2\textbf{w}_1^*\textbf{w}_1 + \tau_1 |\textbf{w}_1^*\bold{H}_{11}\textbf{f}_1|^2}\right)\\&+
\log\left(1 + \frac{\rho_2|\textbf{w}_2^*\bold{H}_{12}\textbf{f}_1 |^2}{\sigma_2^2\textbf{w}_2^*\textbf{w}_2 + \tau_2 |\textbf{w}_2^*\bold{H}_{22}\textbf{f}_2|^2}\right).
\end{split}
\end{equation}   
The goal of this design is to find the optimal precoders and combiners that maximize the objective function (\ref{rate}) under the unit-norm and CA constraints. The optimization problem can be formulated as follows
\begin{equation}\label{maxrate}
\begin{split}
\mathcal{P}: \max\limits_{\bold{f}_u,\bold{w}_v}& \mathcal{I}  
\end{split}
\end{equation}
\begin{equation}\label{c1}
\text{subject to}~~|\bold{f}_1| = |\bold{f}_2| = |\bold{w}_1| = |\bold{w}_2| = 1
\end{equation}
\begin{equation}\label{c2}
~~~~~~~~~~~~~~~~~~~~~~\bold{w}_1,~\bold{w}_2 \in \mathcal{G}^{N_{\text{RX}}}~\text{and}
~\bold{f}_1,~\bold{f}_2 \in \mathcal{G}^{N_{\text{TX}}}
\end{equation}
where (\ref{c1}) is the unit-norm constraint, and (\ref{c2}) is the CA constraint. Note that $\mathcal{G}^M$ is the subspace of the unit-norm CA constraint of dimension $M$ defined by
\begin{equation}
\mathcal{G}^M = \left\{ \boldsymbol{g}\in\mathbb{C}^M~\bigg{|}~g_m = \frac{1}{\sqrt{M}}e^{j\varphi_m},~m=1,\ldots,M  \right\}    
\end{equation}

The projected Gradient Ascent approach is to first evaluate the gradient of the cost function with respect to $\bold{f}_u$, and $\bold{w}_v$ and then project the solutions onto the subspace of the CA constraint. 

Note that the optimization problem is not concave. Under the constraint ($\ref{c1}$), the cost function ($\ref{rate}$) and its gradients are continuous, differentiable and upper bounded. Consequently, the Gradient Ascent of ($\ref{rate}$) is \emph{Lipschitz continuous} with constant $\beta>0$, i.e., for any $x$ and $y$, it exists a non-negative constant $\beta$ such that
\begin{equation}
\|\nabla\mathcal{I}(y) - \nabla\mathcal{I}(x)\|_2 \leq \beta\|y - x\|_2
\end{equation}
According to the following Theorem, we can prove the convergence of the Gradient Ascent as follows
\begin{theorem}
Since the cost function ($\ref{rate}$) is differentiable and its gradients are Lipschitz continuous with a constant $\beta$, then if we run Gradient Ascent for $n$ iterations with a step size (learning rate) $\delta \leq 1/\beta$, it will yield a solution $\mathcal{C}^{(n)}$ satisfying 
\begin{equation}
\mathcal{I}(x^*) - \mathcal{I}(x^{(n)}) \leq \frac{\| x^* - x^{(0)}   \|_2^2}{2\delta n}    
\end{equation}
\end{theorem}
where $\mathcal{I}(x^*)$ is the optimal value. Intuitively, this means that the Gradient Ascent is guaranteed to converge with rate $\mathcal{O}(\frac{1}{n})$. As the gradient is non-zero, the best value for the step size is $1/\beta$ to guarantee that the cost function increases during the optimization cycle. In practice, however, $\beta$ is usually expensive to compute, and this step size is small. To check whether the selected step size efficiently works or not, we can control the evolution of the cost function during the optimization cycle. If the cost function experiences some perturbations, we can adaptively decrease the step size to guarantee the convergence stability.
Differentiating the cost function ($\ref{rate}$) for ($u\neq v$), the gradients are expressed by (\ref{d1}) and (\ref{d2}). 
\begin{equation}{\label{d1}}
\begin{split}
&\nabla_{\bold{w}_v} \mathcal{I}= \frac{\partial \mathcal{I}}{\partial \bold{w}^*_v} = -\frac{\sigma_v^2\textbf{w}_v}{\sigma_v^2\textbf{w}_v^*\textbf{w}_v + \tau_u |\textbf{w}_u\textbf{H}_{uu}\textbf{f}_u|^2}\\&+\frac{\sigma_v^2\textbf{w}_v+\rho_u\textbf{H}_{uv}\textbf{f}_u\textbf{f}_u^*\textbf{H}_{uv}^*\textbf{w}_v}{\sigma_v^2\textbf{w}_v^*\textbf{w}_v^* + \tau_u|\textbf{w}_v^*\textbf{H}_{uv}\textbf{f}_u|^2 + \rho_u|\textbf{w}_v\textbf{H}_{uv}\textbf{f}_u|^2}\\&+\frac{\tau_v\textbf{H}_{vv}\textbf{f}_v\textbf{f}_v^*\textbf{H}_{vv}^*\textbf{w}_v}{\sigma_u^2\textbf{w}_u^*\textbf{w}_u+\tau_v|\textbf{w}_v^*\textbf{H}_{vv}\textbf{f}_v|^2+\rho_v|\textbf{w}_u^*\textbf{H}_{vu}\textbf{f}_v|^2}\\&-\frac{\tau_v\textbf{H}_{vv}\textbf{f}_v^*\textbf{f}_v\textbf{H}_{vv}^*\textbf{w}_v}{\sigma_u^2\textbf{w}_u^*\textbf{w}_u + \tau_v|\textbf{w}_v^*\textbf{H}_{vv}\textbf{f}_v|^2}.
\end{split}
\end{equation}
\begin{equation}{\label{d2}}
\begin{split}
\nabla_{\bold{f}_u} \mathcal{I}=\frac{\partial \mathcal{I}}{\partial \bold{f}^*_u} =&  \frac{\tau_u\textbf{H}_{uu}^*\textbf{w}_u\textbf{w}_u^*\textbf{H}_{uu}\textbf{f}_u + \rho_u\textbf{H}_{uv}^*\textbf{w}_v\textbf{w}_v^*\textbf{H}_{uv}\textbf{f}_u}{\sigma_v^2\textbf{w}_v^*\textbf{w}_v + \tau_u|\textbf{w}_u^*\textbf{H}_{uu}\textbf{f}_u|^2+\rho_u|\textbf{w}_v^*\textbf{H}_{uv}\textbf{f}_u|^2}\\&-\frac{\tau_u\textbf{H}_{uu}^*\textbf{w}_u\textbf{w}_u^*\textbf{H}_{uu}\textbf{f}_u}{\sigma_v^2\textbf{w}_v^*\textbf{w}_v + \tau_v|\textbf{w}_v^*\textbf{H}_{vv}\textbf{f}_v|^2}.
\end{split}    
\end{equation}
The gradient search algorithm is given by
 \begin{algorithm}[H]
 \caption{Adaptive Gradient Ascent}\label{algo}
 \begin{algorithmic}[1]
 \renewcommand{\algorithmicrequire}{\textbf{Input:}}
 \renewcommand{\algorithmicensure}{\textbf{Output:}}
 \renewcommand{\algorithmiccomment}{$\triangleright~$}
 \REQUIRE $\delta$, $\epsilon$, $\bold{H}_{uv},~\bold{H}_{vu}$,~$\bold{H}_{vv}$,~$\bold{H}_{uu}$  ($u,~v \in \{0,~1\},~u \neq v$)
 \ENSURE $\bold{w}^{\star}_v$,~$\bold{f}^{\star}_u$\\
 %\State \textbf{Initialize}$\boldsymbol{z}_0 = \boldsymbol{a}$
 %\\ \textit{Initialisation} :
  \STATE \textbf{Initialize} $\bold{w}_v = \bold{w}^{(0)}_v,~\bold{f}_u = \bold{f}^{(0)}_u$
  \WHILE{$|\mathcal{I}^{(n+1)} - \mathcal{I}^{(n)}| > \epsilon$} 
  \STATE $\bold{w}_v^{(n+1)} \gets \bold{w}_v^{(n)}+\delta \nabla_{\bold{w}_v}\mathcal{I}^{(n)}$
  \STATE $\bold{w}_v^{(n+1)} \gets \frac{\bold{w}_v^{(n+1)}}{\|\bold{w}_v^{(n+1)}\|_2}$
  \COMMENT{Unit-norm constraint}
  \STATE $\bold{w}_v^{(n+1)} \gets \frac{\bold{w}_v^{(n+1)}}{\sqrt{N_{r,v}}\left|\bold{w}_v^{(n+1)}\right|}$
  \COMMENT{CA constraint}
  \STATE $\bold{f}_u^{(n+1)} \gets \bold{f}_u^{(n)}+\delta \nabla_{\bold{f}_u}\mathcal{I}^{(n)}$
  \STATE $\bold{f}_u^{(n+1)} \gets \frac{\bold{f}_u^{(n+1)}}{\|\bold{f}_u^{(n+1)}\|_2}$
   \STATE $\bold{f}_u^{(n+1)} \gets \frac{\bold{f}_u^{(n+1)}}{\sqrt{N_{t,u}}\left|\bold{f}_u^{(n+1)}\right|}$
    %\IF{$\mathcal{C}^{(n+1)}<\mathcal{C}^{(n)$}
    %\STATE Adapt $\delta$
    %\ENDIF
    \IF{$\mathcal{I}^{(n+1)}<\mathcal{I}^{(n)}$} \STATE {Adapt $\delta$} \ENDIF
  %\UNTIL{convergence}
  \ENDWHILE
 \RETURN $\bold{w}_v^\star,~\bold{f}_u^\star$ 
 \end{algorithmic} 
 \end{algorithm}
Note that the objective of designing the optimal beamformers is to maximize the sum rate, i.e., maximizing the numerator and minimizing the denominator. This is perfectly translated into seeking for the best beam pair between TX and RX (maximize $|\bold{w}_v^*\bold{H}_{uv}\bold{f}_u|^2$) and assigning null weights to the interference direction (minimize $|\bold{w}_u^*\bold{H}_{uu}\bold{f}_u|^2$).

We also note that the Gradient Ascent convergence is also sensitive to the choice of the initialization as it affects the number of iterations required for the convergence. The best initialization is the one that achieves the convergence with the minimum number of iterations. Basically, we will consider the following initialization vectors for the precoder $\bold{f}_u$ and combiner $\bold{w}_v$.
\begin{enumerate}
    \item Normalized complex Gaussian vectors.
    %\item The most right and left singular vectors of the singular value decomposition (SVD) of the channel $\bold{H}_{uv}(\bold{H}_{vv}\bold{H}_{vv}^*)^{-1}$.
    \item The most right and left singular vectors of the singular value decomposition (SVD) of the channel $\bold{H}_{uv}$.
\end{enumerate}

To characterize the system performance, we further derive the upper bound of the achievable rate. Neglecting the interference ($\textbf{H}_{\text{SI}}=0$) and applying the SVD on the channels $\textbf{H}_{12}$ and $\textbf{H}_{21}$, the upper bound is given by 
\begin{equation}\label{upper}
\mathcal{I}_{\text{ub}} = \log\left(1 + \lambda^2_{12}\frac{\rho_1}{\sigma_2^2}\right) +  \log\left(1 + \lambda^2_{21}\frac{\rho_2}{\sigma_1^2}\right)  
\end{equation}
where $\lambda_{12}$, and $\lambda_{21}$ are the maximum singular values of $\bold{H}_{12}$, and $\bold{H}_{21}$, respectively. Although, the rate (\ref{upper}) is not achievable, it serves as a benchmarking tool to quantify the system performance.

%% file: relay.tex
\section{Application of Dual-Hop Full-Duplex Relaying Network}
\begin{figure}[H]
    \centering
    \includegraphics[scale=.49]{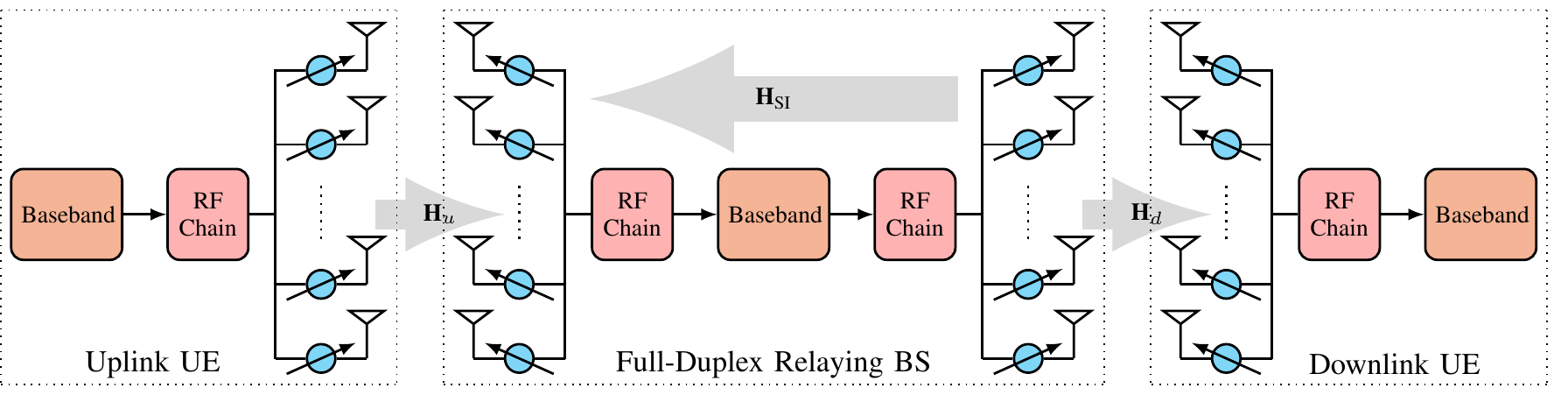}
    \caption{Dual-hop FD relaying system. The TX and RX half-duplex nodes are the uplink and downlink users equipments (UEs) while the relay is considered as a full-duplex base station (BS). The uplink and downlink data of the UEs are independent.}
\label{relay}
\end{figure}
Fig.~\ref{relay} illustrates the transmit and receive nodes communicating through an intermediary relay which operates in FD mode. The SI leakage from the TX to RX parts of the relay is modeled by the channel matrix $\bold{H}_{\text{SI}}$ given by (\ref{channelSI}). The received uplink signal at the relay is given by
\begin{equation}\label{uplinksignal}
y_u = \sqrt{\rho_u}\textbf{w}_r^*\textbf{H}_u\textbf{f}_us_u + \sqrt{\tau_u} \textbf{w}_r^*\textbf{H}_{\text{SI}}\textbf{f}_r s_d + \textbf{w}_r^*\textbf{n}_r    
\end{equation}
where $\bold{H}_u$ is the uplink channel matrix, $\bold{f}_u$ is the analog precoder of UE, $\bold{w}_r$ and $\bold{f}_r$ are the combiner and precoder at the relay, $\bold{s}_u$ and $\bold{s}_d$ are the transmitted symbols for uplink and downlink, respectively. Also $\rho_u$ is the transmit power at the UE, $\tau_u$ is the SI power and $\bold{n}_r$ is the noise at the relay.

For downlink scenario, the received signal $y_d$ is given by
\begin{equation}\label{downlinksignal}
y_d = \sqrt{\rho_d}\bold{w}^*_d\bold{H}_d\bold{f}_r s_d + \bold{w}_d^*\bold{n}_d
\end{equation}
where $\bold{w}_d$ is the analog combiner of the UE, $\rho_d$ is the average transmit power at the relay and $\bold{n}_d$ is the noise at the receive UE.
\subsection{Spectral Efficiency}
Capitalizing on the uplink and downlink signals formulation, the sum rate is expressed by
\begin{equation}\label{ratedual}
\begin{split}
\mathcal{I} =& \overbrace{\log\left( 1 + \frac{\rho_u|\bold{w}_r^*\bold{H}_u\bold{f}_u|^2}{ \tau_u|\bold{w}_r^*\bold{H}_{\text{SI}}\bold{f}_r|^2 + \sigma^2\bold{w}_r^*\bold{w}_r}  \right)}^{\textsf{Uplink}} \\&+ \underbrace{\log\left(1 + \frac{\rho_d|\textbf{w}_d^*\textbf{H}_d\textbf{f}_r|^2}{\sigma^2\bold{w}_d^*\bold{w}_d}\right)}_{\textsf{Downlink}}. 
\end{split}
\end{equation}

The goal is to design the optimal precoders and combiners to maximize the spectral efficiency (\ref{ratedual}). Given that (\ref{ratedual}) has the same generic form as (\ref{rate}), the derivation of the gradients with respect to $\bold{f}_u$, $\bold{f}_r$, $\bold{w}_r$ and $\bold{w}_d$ follow the same rules
as the analysis of the two-nodes FD network.
\subsection{Outage Probability}
Once a transmission strategy is specified, the corresponding outage probability for rate $R$ (bits/s/Hz) is then
\begin{equation}
    P_{\textsf{out}}(\textsf{\scriptsize{SNR}}, R) = \mathbb{P}[\mathcal{I}(\textsf{\scriptsize{SNR}})<R].
\end{equation}

%% file: numerical.tex
\section{Numerical Results}
\begin{table}[H]
\renewcommand{\arraystretch}{1}
\caption{System Parameters}
\label{param}
\centering
\begin{tabular}{|l|l|}
\hline\hline
%\bfseries Parameter & \bfseries Symbol & \bfseries Value\\
Carrier frequency & 28 GHz\\
\hline
Bandwidth &850 MHz\\
\hline
Number of transmit antennas& 16\\
\hline
Number of receive antennas & 16\\
\hline
Antenna separation & $\frac{\lambda}{2}$\\
\hline
Antenna correlation & None\\
\hline
Number of clusters & 6\\
\hline 
Number of rays per cluster & 8\\
\hline
Angular spread& 20$^{\circ}$\\
\hline
Distance between FD node arrays& 2$\lambda$\\
\hline
Angle between FD node arrays & $\frac{\pi}{6}$\\
\hline
Self-interference power & 0 dB\\
\hline
Rician factor& 5 dB\\
\hline
Step size ($\delta$) & 1\\
\hline
Convergence criterion ($\epsilon$)& 1e-5\\
\hline\hline
\end{tabular}
\end{table}
In this section, we will provide the performance of the two-nodes FD network and the dual-hop FD relaying system following the discussion of the results. A practical choice of the learning rate is to start with $\delta = 1$ and whenever the cost function experiences any possible perturbation during the optimization cycle, we adaptively divide the step size by half to guarantee the convergence stability. In addition, we run 1000 loops of Monte Carlo simulation to average the performance results. Unless otherwise stated, TABLE \ref{param} summarizes the values of the system parameters.
\subsection{Number of Antennas and Signal-to-Interference Ratio (SIR) }
\begin{figure}[H]
    \centering
    \includegraphics{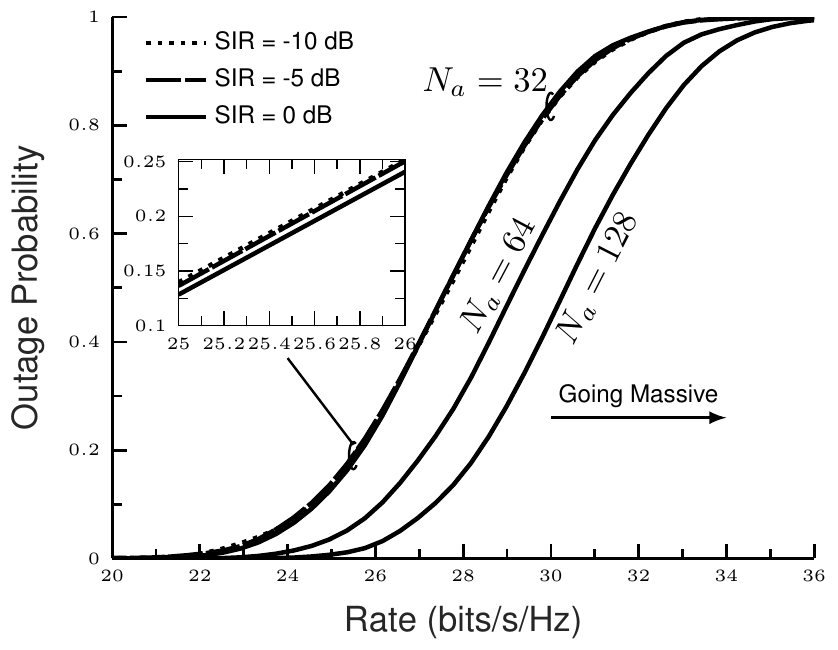}
    \caption{Outage probability of the sum rate performance for the relaying system: Evaluations are performed for various {$\tt SIR$} levels and for different number of active antennas at the relay $N_a$. Note that the number of antennas at UE is 2x1.}
    \label{pict6}
\end{figure}

Fig. \ref{pict6} illustrates the outage performance with respect to a given range of target rate requirements. We observe that the FD relaying system can support higher target rate requirements. For different {$\tt SIR$} levels, we notice that the systems performances are roughly unchanged and this proves the resiliency of the proposed technique against the interference, i.e., the gradient search completely eliminates the interference for the suggested SI power values. Moreover, the massive number of antennas can provide further degree of freedom not only to cancel the interference but also to provide spatial multiplexing gain to improve the achievable rate and support higher target rate requirements.
\subsection{Gradient Search Vs. Conventional Approaches}
\begin{figure}[H]
    \centering
    \includegraphics{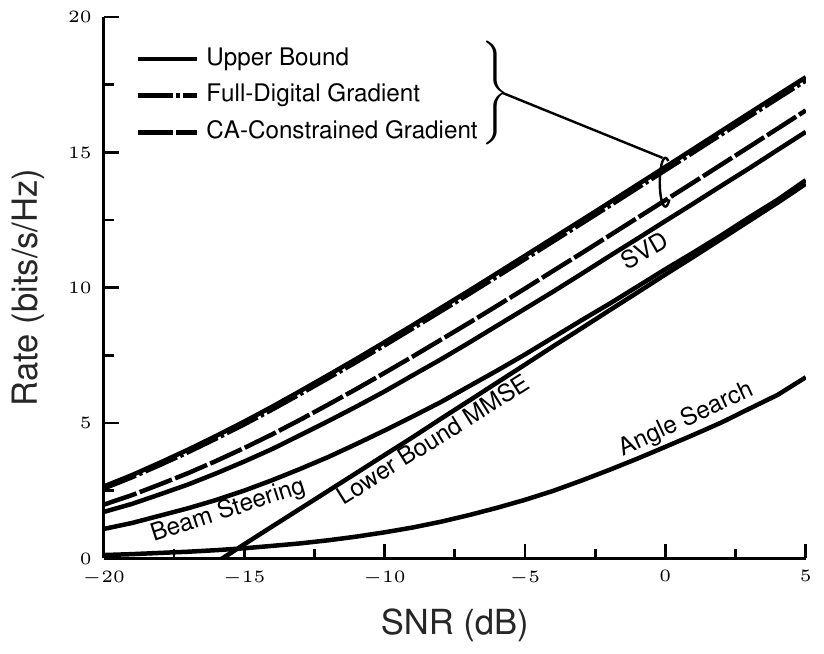}
    \caption{Sum rate performance for the two-nodes FD system: Comparisons are made between the gradient search approach and conventional interference cancellation techniques. Full-digital as well as CA constrained gradient are presented while a benchmarking upper bound is provided to quantify the limitations of the proposed technique.}
    \label{pict4}
\end{figure}
Fig.~\ref{pict4} illustrates the achievable sum rate performance of the two-nodes FD network for various approaches with respect to the average {$\tt SNR$}. Clearly, we observe the detrimental effect of the CA constraint on the performance presented by \cite{14}. It is widely known that the SVD beamforming approach is optimal since it diagonalizes the channels, however, the SVD does not account for the interference in this case which results in a rate loss of roughly 3 bits/s/Hz compared to the upper bound at 5 dB of {$\tt SNR$}. Around an average {$\tt SNR$} of 5 dB, the rate for full-digital design (only unit-norm constraint) is roughly 18 bits/s/Hz and it completely coincides with the upper bound. In other terms, the unconstrained gradient completely eliminates the interference. Besides, the rate constrained by CA is 17 bits/s/Hz, resulting in a loss about 0.5 bits/s/Hz. Inversely and for a given rate of 17 bits/s/Hz, the proposed constrained approach requires around 2 dB of {$\tt SNR$} to compensate for the CA loss and achieves the upper bound. We conclude that our approach substantially mitigates the losses incurred by the CA constraint compared to the other approaches and it requires an {$\tt SNR$} of 2 dB to completely compensate for the CA losses which is commonly acceptable for practical design.
%\vspace*{-.3cm}
%\vspace*{-.4cm}
\subsection{Duplex Modes}
\begin{figure}[H]
    \centering
    \includegraphics{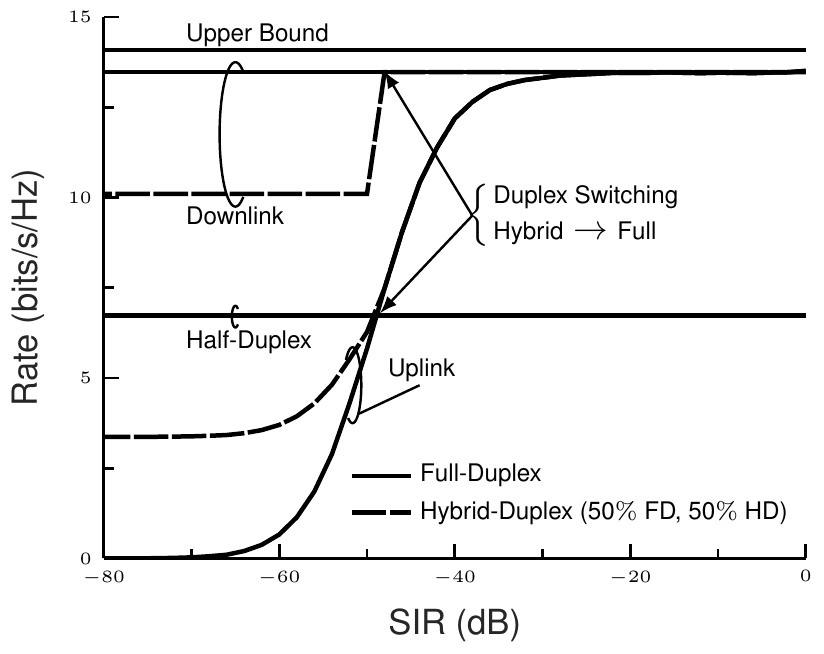}
    \caption{Rate performance results for relaying systems: Comparisons are made for CA constrained gradient. First comparison takes place between uplink, downlink in FD mode, HD and hybrid-duplex modes. For the later duplex mode, the time duration is equally divided into FD and HD operations at the BS. Note that the {$\tt SIR$} is changing by fixing the transmit power ({$\tt SIR$} = 10 dB) and varying the SI power. Note that the number of antennas at the relay and UE are 4x4 and 2x1, respectively.}
    \label{pict7}
\end{figure}
Fig. \ref{pict7} illustrates the performance of the relaying system in terms of the achievable rate with respect to the {$\tt SIR$} and for the CA constrained gradient search. Since the downlink FD and HD modes are interference-free, the corresponding rates are constant since the {$\tt SNR$} is fixed at 10 dB. For a low {$\tt SIR$} range from -80 to -40 dB which corresponds to a cell-edge user, the uplink rate is substantially degraded resulting in a practical null rate. Within the same range, the BS can operate in HD mode to avoid the severe interference and offers the uplink UE with relatively acceptable rate. However, this approach will also decrease the downlink rate from 14.5 to around 10 bits/s/Hz. 
A practical solution can be applied by introducing a new operating scheme called hybrid-duplex to establish a tradeoff between the uplink and downlink rates. In this case, the uplink cell-edge user still achieves an acceptable rate around 3.75 bits/s/Hz instead of HD (7.23 bits/s/Hz) but this duplex mode still offers a better downlink rate roughly 10 bits/s/Hz. Although hybrid-duplex improves the uplink cell-edge user at the expense of the downlink rate, the downlink UE still achieves better rate compared to HD mode. Starting from an {$\tt SIR$} of -50 dB and up to 0 dB (for middle and near users), it is recommended to switch from hybrid-duplex operation to FD mode. These remarks lead to think about how to further improve the uplink and downlink spectral efficiency for hybrid-duplex mode. In other terms, we need to dedicate a careful attention on how to design two optimal switching points that have to be primarily adaptive to the {$\tt SIR$} levels to maximize the uplink and downlink ergodic spectral efficiency. The first switching occurs within the hybrid-duplex mode, i.e., how to optimally allocate the time fractions for FD and HD, while the second switching occurs between hybrid and FD modes. For now, we defer the design of these optimal switching points as a future extension for this work.

\subsection{Gradient Initializations}
In the sequel, we focalize on the impacts of the gradient intializations on the number of iterations required for the convergence. We quantify the uplink rate metric for the FD relaying systems under CA constraint evaluated for different levels of {$\tt SNR$} and {$\tt SIR$} for cell-edge and near users. To evaluate the efficiency of the proposed technique, we further define a new metric called the rate gain to provide insights into the FD performance compared to the HD mode.
 \begin{equation}
 \Gamma[\%] = \frac{ \mathcal{I}^{\text{FD}}(\scriptsize{\textsf{SNR}) } - \mathcal{I}^{\text{HD}}(\scriptsize{\textsf{SNR}) } }{ \mathcal{I}^{\text{HD}}(\scriptsize{\textsf{SNR}) } } \times 100.   
 \end{equation}
TABLE \ref{gain-loss} provides the uplink rate gain of the FD relaying system relative to HD mode as well as the number of iterations required for the convergence and for a couple of beamforming initializations. We note that the algorithm always converges regardless of the initial condition which means that the adaptive step size works efficiently.
 \begin{table}[H]
\renewcommand{\arraystretch}{1.0}
\caption{Uplink Rate Gain and Complexity}
\label{gain-loss}
\centering
\begin{tabular}{|l|c|c|c|}
\hline\hline
\multicolumn{4}{|c|}{ \scriptsize{\sffamily{SIR} = - 60 dB (cell-edge user)}  }\\ \hline
{\scriptsize{\sffamily{SNR}}} (dB) & -30 & 0 & 10\\\hline
{\scriptsize{\sffamily{Gain}}} [\%] & -99.998 & -99.986 & -96.77\\\hline
{\scriptsize{\sffamily{SVD}}} & 25 & 27 & 36\\\hline
{\scriptsize{\sffamily{Gaussian}}} & 25 & 32 & 48\\\hline
\multicolumn{4}{|c|}{ \scriptsize{\sffamily{SIR} = 0 dB (near user)}  }\\ \hline
{\scriptsize{\sffamily{SNR}}} (dB) & -30 & 0 & 10\\\hline
{\scriptsize{\sffamily{Gain}}} [\%] & 99.83 & 101.31 & 101.276\\\hline
{\scriptsize{\sffamily{SVD}}} & 60 & 62 & 63\\\hline
{\scriptsize{\sffamily{Gaussian}}} & 70 & 72 & 75\\\hline\hline
    \end{tabular}
 \end{table} 
 
 Most importantly, the cost function (spectral efficiency) converges rapidly for the SVD approach since the initialized vectors implicitly acquire the prior knowledge of the channels which consequently speeds up the convergence. The Gaussian initialization requires more iterations to converge since it is blindly established without accounting for the channels information. We further observe that the FD system performance is severely degraded for the uplink cell-edge user which is exposed to the high SI power (60 dB). In agreement to the results illustrated by Fig. \ref{pict7}, the proposed algorithm cannot reduce the SI power (around 60 dB) to some acceptable floor and result in absolute null spectral efficiency even for low and high {$\tt SNR$}. For near user, the algorithm achieves a high SI power reduction and realizes a gain of around 99$\%$. Besides, the proposed technique enables the FD system to provide the uplink user with more than 2 times the rate achieved by HD for 0 and 10 dB of {$\tt SNR$}. In addition, we note that for near user, the convergence requires more iterations to maximize the rate gain and reduce the SI power below the noise floor. Unlike the near user, the analog-to-digital converter (ADC) is pronouncedly saturated by the high SI power (cell-edge user). Thereby, the cost function is bounded by an irreducible floor created by the SI power and the algorithm converges at early stage in few iterations without maintaining the interference below the noise floor.

%% file: conclusion.tex
\section{Conclusion}
In this work, we proposed the adaptive Gradient Ascent to maximize the sum rate and minimizing the losses incurred by the CA constraint. Thanks to the adaptive step size, the results show that the proposed algorithm always converges regardless of the initial condition. We also prove that the number of iterations can be substantially reduced when the initial condition accounts for the knowledge of the channels. Furthermore, the results show the efficiency of our approach in comparison with HD mode and conventional beamforming designs implemented for FD systems. We observed that the proposed technique minimizes the losses introduced by the CA constraint (around 0.5 bits/s/Hz) and the system requires roughly 2 dB of {$\tt SNR$} to compensate for such a loss. Moreover, our approach seems to work much better for massive MIMO setting where substantial degree of freedom are provided for SI cancellation and spatial multiplexing gain resulting in relatively interference-free MIMO FD relaying system. In practice, the hardware impairments such as the power amplifier nonlinearities, phase noise, active/passive intermodulation, in phase and quadrature imbalance and channel estimation error have to accounted as additional sources of SI and most likely the SI channel  is unknown. As a future direction, we will consider a channel estimation process of the aggregate interference and imperfection channel and we will implement a modified version of our approach to compensate for the inaccuracy of the channel estimation and the CA constraint to get an achievable rate with tolerant losses. In addition, we are planning to design the optimal switching points for the hybrid-duplex mode which deserve a careful attention.
%for our future work.
%design the hybrid-duplex mode which triggers us to think about how to design the optimal switching points 

%% file: main.bbl
% Generated by IEEEtran.bst, version: 1.13 (2008/09/30)
\begin{thebibliography}{10}
\providecommand{\url}[1]{#1}
\csname url@samestyle\endcsname
\providecommand{\newblock}{\relax}
\providecommand{\bibinfo}[2]{#2}
\providecommand{\BIBentrySTDinterwordspacing}{\spaceskip=0pt\relax}
\providecommand{\BIBentryALTinterwordstretchfactor}{4}
\providecommand{\BIBentryALTinterwordspacing}{\spaceskip=\fontdimen2\font plus
\BIBentryALTinterwordstretchfactor\fontdimen3\font minus
  \fontdimen4\font\relax}
\providecommand{\BIBforeignlanguage}[2]{{%
\expandafter\ifx\csname l@#1\endcsname\relax
\typeout{** WARNING: IEEEtran.bst: No hyphenation pattern has been}%
\typeout{** loaded for the language `#1'. Using the pattern for}%
\typeout{** the default language instead.}%
\else
\language=\csname l@#1\endcsname
\fi
#2}}
\providecommand{\BIBdecl}{\relax}
\BIBdecl

\bibitem{e1}
E.~{Balti} and M.~{Guizani}, ``{Impact of Non-Linear High-Power Amplifiers on
  Cooperative Relaying Systems},'' \emph{IEEE Transactions on Communications},
  vol.~65, no.~10, pp. 4163--4175, Oct 2017.

\bibitem{e2}
E.~{Balti}, M.~{Guizani}, B.~{Hamdaoui}, and B.~{Khalfi}, ``{Aggregate Hardware
  Impairments Over Mixed RF/FSO Relaying Systems With Outdated CSI},''
  \emph{IEEE Transactions on Communications}, vol.~66, no.~3, pp. 1110--1123,
  March 2018.

\bibitem{e3}
E.~{Balti} and M.~{Guizani}, ``{Mixed RF/FSO Cooperative Relaying Systems With
  Co-Channel Interference},'' \emph{IEEE Transactions on Communications},
  vol.~66, no.~9, pp. 4014--4027, Sep. 2018.

\bibitem{e4}
E.~{Balti}, M.~{Guizani}, B.~{Hamdaoui}, and Y.~{Maalej}, ``{Partial Relay
  Selection for Hybrid RF/FSO Systems with Hardware Impairments},'' in
  \emph{2016 IEEE Global Communications Conference (GLOBECOM)}, Dec 2016, pp.
  1--6.

\bibitem{e5}
E.~{Balti}, M.~{Guizani}, and B.~{Hamdaoui}, ``{Hybrid Rayleigh and
  Double-Weibull over impaired RF/FSO system with outdated CSI},'' in
  \emph{2017 IEEE International Conference on Communications (ICC)}, May 2017,
  pp. 1--6.

\bibitem{e6}
E.~{Balti}, M.~{Guizani}, B.~{Hamdaoui}, and B.~{Khalfi}, ``{Mixed RF/FSO
  Relaying Systems with Hardware Impairments},'' in \emph{GLOBECOM 2017 - 2017
  IEEE Global Communications Conference}, Dec 2017, pp. 1--6.

\bibitem{e7}
E.~{Balti} and B.~K. {Johnson}, ``{Tractable Approach to MmWaves Cellular
  Analysis with FSO Backhauling under Feedback Delay and Hardware
  Limitations},'' \emph{IEEE Transactions on Wireless Communications}, pp.
  1--1, 2019.

\bibitem{eT}
E.~Balti, ``\BIBforeignlanguage{English}{Analysis of hybrid free space optics
  and radio frequency cooperative relaying systems},'' Master's thesis, 2018.

\bibitem{asym}
E.~Balti and B.~K. Johnson, ``Asymmetric rf/fso relaying with hpa
  non-linearities and feedback delay constraints,'' 2019.

\bibitem{maalej}
Y.~{Maalej}, A.~{Abderrahim}, M.~{Guizani}, B.~{Hamdaoui}, and E.~{Balti},
  ``Advanced activity-aware multi-channel operations1609.4 in vanets for
  vehicular clouds,'' in \emph{2016 IEEE Global Communications Conference
  (GLOBECOM)}, 2016, pp. 1--6.

\bibitem{neji1}
N.~{Mensi}, M.~{Guizani}, and A.~{Makhlouf}, ``Study of vehicular cloud during
  traffic congestion,'' in \emph{2016 4th International Conference on Control
  Engineering Information Technology (CEIT)}, 2016, pp. 1--6.

\bibitem{neji2}
N.~{Mensi}, A.~{Makhlouf}, and M.~{Guizani}, ``Incentives for safe driving in
  vanet,'' in \emph{2016 4th International Conference on Control Engineering
  Information Technology (CEIT)}, 2016, pp. 1--6.

\bibitem{surv}
Y.~Niu, Y.~Li, D.~Jin, L.~Su, and A.~V. Vasilakos, ``{A survey of millimeter
  wave communications (mmWave) for 5G: opportunities and challenges},''
  \emph{Wireless Networks}, vol.~21, no.~8, pp. 2657--2676, Nov 2015.

\bibitem{rheath}
R.~W. {Heath}, N.~{González-Prelcic}, S.~{Rangan}, W.~{Roh}, and A.~M.
  {Sayeed}, ``{An Overview of Signal Processing Techniques for Millimeter Wave
  MIMO Systems},'' \emph{IEEE Journal of Selected Topics in Signal Processing},
  vol.~10, no.~3, pp. 436--453, April 2016.

\bibitem{5gnr}
E.~{Onggosanusi}, M.~S. {Rahman}, L.~{Guo}, Y.~{Kwak}, H.~{Noh}, Y.~{Kim},
  S.~{Faxer}, M.~{Harrison}, M.~{Frenne}, S.~{Grant}, R.~{Chen}, R.~{Tamrakar},
  and a.~Q.~{Gao}, ``{Modular and High-Resolution Channel State Information and
  Beam Management for 5G New Radio},'' \emph{IEEE Communications Magazine},
  vol.~56, no.~3, pp. 48--55, March 2018.

\bibitem{v2v}
J.~{Choi}, V.~{Va}, N.~{Gonzalez-Prelcic}, R.~{Daniels}, C.~R. {Bhat}, and
  R.~W. {Heath}, ``Millimeter-wave vehicular communication to support massive
  automotive sensing,'' \emph{IEEE Communications Magazine}, vol.~54, no.~12,
  pp. 160--167, 2016.

\bibitem{sub6}
E.~Balti and B.~K. Johnson, ``Sub-6 ghz microstrip antenna: Design and
  radiation modeling,'' 2019.

\bibitem{isol}
A.~Kiayani, M.~Z. Waheed, L.~Anttila, M.~Abdelaziz, D.~Korpi, V.~Syrjala,
  M.~Kosunen, K.~Stadius, J.~Ryynanen, and M.~Valkama, ``Adaptive nonlinear rf
  cancellation for improved isolation in simultaneous transmit–receive
  systems,'' \emph{IEEE Transactions on Microwave Theory and Techniques},
  vol.~66, no.~5, p. 2299–2312, May 2018.

\bibitem{19}
B.~P. {Day}, A.~R. {Margetts}, D.~W. {Bliss}, and P.~{Schniter}, ``Full-duplex
  bidirectional mimo: Achievable rates under limited dynamic range,''
  \emph{IEEE Transactions on Signal Processing}, vol.~60, no.~7, pp.
  3702--3713, July 2012.

\bibitem{analog}
J.~I. Choi, M.~Jain, K.~Srinivasan, P.~Levis, and S.~Katti, ``Achieving single
  channel, full duplex wireless communication,'' in \emph{Proceedings of the
  Sixteenth Annual International Conference on Mobile Computing and
  Networking}, ser. MobiCom '10.\hskip 1em plus 0.5em minus 0.4em\relax New
  York, NY, USA: Association for Computing Machinery, 2010, p. 1–12.

\bibitem{digital}
Q.~{Wang}, Y.~{Dong}, X.~{Xu}, and X.~{Tao}, ``Outage probability of
  full-duplex af relaying with processing delay and residual
  self-interference,'' \emph{IEEE Communications Letters}, vol.~19, no.~5, pp.
  783--786, 2015.

\bibitem{14}
Z.~{Xiao}, P.~{Xia}, and X.~{Xia}, ``{Full-Duplex Millimeter-Wave
  Communication},'' \emph{IEEE Wireless Communications}, vol.~24, no.~6, pp.
  136--143, Dec 2017.

\bibitem{unconst}
X.~Liu, Z.~Xiao, L.~Bai, J.~Choi, P.~Xia, and X.-G. Xia, ``{Beamforming Based
  Full-Duplex for Millimeter-Wave Communication},'' \emph{Sensors}, vol.~16,
  no.~7, 2016.

\bibitem{balti2020modified}
E.~Balti, N.~Mensi, and S.~Yan, ``A modified zero-forcing max-power design for
  hybrid beamforming full-duplex systems,'' 2020.

\end{thebibliography}
